\documentclass[%
superscriptaddress,
reprint, nature,
 amsmath,amssymb,
floatfix,
]{revtex4-1}

\usepackage{mathcomp}
\usepackage{graphicx}
\usepackage{dcolumn}
\usepackage{bm}
\usepackage{txfonts}  
\usepackage{makecell} 
\usepackage[utf8]{inputenc}
\usepackage{dirtytalk}

\usepackage[usenames,dvipsnames,svgnames,table]{xcolor}

\newcommand{\lrm}[1]{_{\text{#1}}}
\newcommand{\hrm}[1]{^{\text{\tiny{#1}}}}

\usepackage{hyperref}
\hypersetup{
    colorlinks=true,
    linkcolor=blue,
    urlcolor=blue,
    citecolor = blue
    }

\begin{document}

\preprint{}

\title{Molecular motors enhance microtubule lattice plasticity}

\author{William Lecompte}
\affiliation{Université Grenoble Alpes / CNRS, LIPhy, 38000 Grenoble, France}%

%
%

\author{Karin John}%
 \email{karin.john@univ-grenoble-alpes.fr}
\affiliation{Université Grenoble Alpes / CNRS, LIPhy, 38000 Grenoble, France}%




\date{\today}

\begin{abstract}

Microtubules are key structural elements of living cells that are crucial for cell division, intracellular transport and motility.
Recent experiments have shown that microtubule severing proteins  and molecular motors stimulate the direct and localized incorporation of free tubulin into the shaft. 
%
However, a mechanistic picture how microtubule associated proteins affect the lattice is completely missing. 
Here we theoretically explore a potential mechanism of lattice turnover stimulated by processive molecular motors in which a weak transient destabilization of the lattice by the motor stepping promotes the formation of mobile vacancies.
In the absence of free tubulin the defect rapidly propagates leading to a complete fracture. In the presence of free tubulin, the motor walk induces a vacancy drift in the direction opposite of the motor walk. The drift is accompanied by the direct and localized incorporation of free tubulin along the trajectory of the vacancy. Our results are consistent with experiments and strongly suggest that a weak lattice-motor interaction is responsible for an augmented microtubule shaft plasticity.

\end{abstract}

\maketitle


\section*{Introduction}

Microtubules (MTs) are self-organized polar tube-like polymers and constitute a major component of the cytoskeleton. They play a central role in cell division, intracellular transport, and motility. 
MTs are dynamic dissipative structures, which grow or shrink primarily by tubulin dimer addition or removal at their extremities, labelled (+) and (-)-end. Their non-equilibrium behavior results from the irreversible hydrolysis of GTP-tubulin into GDP-tubulin upon polymerization \cite{carlier1982guanosine} and manifests itself as stochastic transitions between growth and shrinkage phases, called dynamic instability \cite{mitchison1984dynamic,walker1988dynamic,howard2003dynamics,duellberg2016size,aher2018tipping}.
%
%
The dynamic instability of the MT tip has been a major focus of MT research over the past 30 years. In contrast the dynamics of the MT shaft has been considered as an inert structure, due to the high stability of the intact lattice far away from the extremities \footnote{Estimates of the change in free energy upon transferring a dimer from the fully occupied lattice into the surrounding medium range from 35 to 80 $k_\mathrm{B}T$ per dimer \cite{vanburen2002estimates,vanburen2005mechanochemical,sept2003physical}.}.
However, an early experiment with end-stabilized MTs by Dye et al. \cite{dye1992endstabilized} clearly showed  that the shaft may loose and incorporate tubulin dimers directly. 
Later it was shown that GTP dimers (or dimers in the GTP conformation) exist outside of the cap region \cite{dimitrov2008detection}, without a clear picture, how the GTP-state could survive sufficiently long to be detectable in the shaft.
A very recent series of experiments revealed that the shaft lattice exhibits a spontaneous dynamics, part of which is linked to lattice dislocations \cite{reid2017manipulation,schaedel2019lattice}. Perturbing the lattice externally via periodic weakly mechanical forcing \cite{schaedel2015microtubules} or the activity of MT associated proteins, e.g. MT servering enzyms or molecular motors \cite{vemu2018severing,triclin2021self,budaitis2022kinesin,andreu2022motor} has been shown to facilitate the localized incorporation of tubulin dimers from the surrounding medium into the lattice. Futhermore, regions of high curvature or regions in close contact either with a surface or another microtubule could be sites of direct tubulin exchange \cite{deforges2016localized}.

%

It has been shown experimentally and theoretically that dislocations are preferential sites of lattice dynamics \cite{schaedel2019lattice}. These structures are inherent in the lattice and are created during the polymerization process. The same holds for recently identified multiseam MTs \cite{guyomar2021structural}, which entail the existence of point defects of the size of a tubulin monomer. However, the experimentally observed increase in shaft plasticity due to severing enzymes and molecular motors \cite{vemu2018severing,triclin2021self,budaitis2022kinesin,andreu2022motor} suggests the nucleation of defects in the intact lattice.
One obvious possibility of {\it de novo} created sites of lattice exchange are vacancies of the size of a single tubulin dimer. These point defects have been identified by scanning force microscopy of dynamic MTs \cite{schaap2004resolving} and it has already been speculated that long-lived point defects may serve as point of attack for MT severing enzymes \cite{davis2002importance}. A recent experiment suggests that the combined walk of several kinesins is able remove tubulin dimers from the lattice \cite{kuo2022force}.

Here, we explore theoretically the role of point defects in the MT shaft plasticity in the presence of molecular motors. Our objective is to provide a first mechanistic concept consistent with recent experiments on MT-motor interactions \cite{triclin2021self,budaitis2022kinesin,andreu2022motor}.
To that end we employ a kinetic Monte Carlo model and investigate the (i) kinetics of the formation of point defects, (ii) the progression of the defect size until complete MT fracture in the absence of free tubulin, and (iii) the dynamics of point defects in the presence of free tubulin. These processes will be studied in the absence and presence of processive molecular motors, which transiently and locally weakly destabilize the lattice as they walk along the lattice.
Since many details of the dynamic properties of the MT shaft lattice are unknown, we are focusing on very basic processes to recover relevant length and time scale of the MT shaft dynamics that were observed experimentally.

\section*{Model}

\label{sec:model}

\begin{figure*}[ht]
    \centering
    \includegraphics[width=0.9\linewidth]{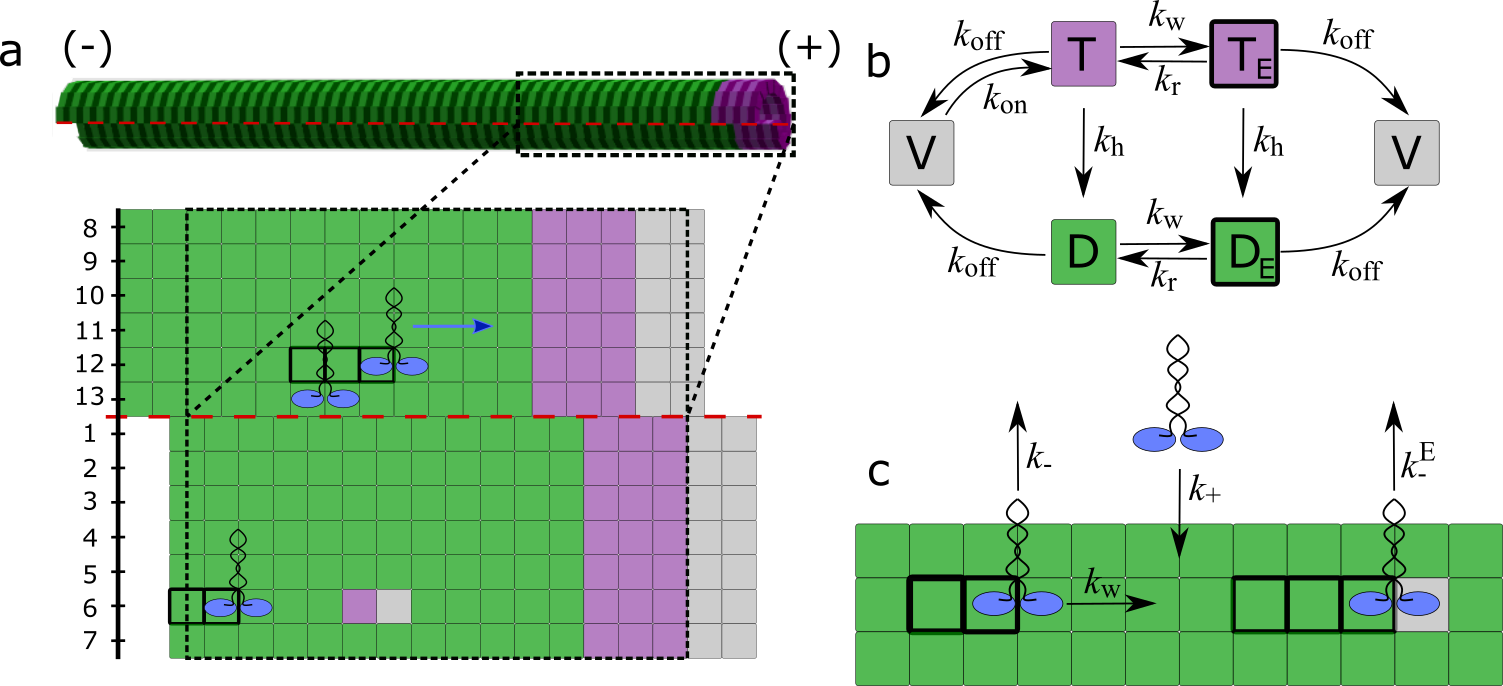}
    \caption{Schematic setup of the kinetic Monte Carlo model for the MT lattice and processive (+)-end directed molecular motors. {\bf a:} The canonical  lattice configuration with (+)-end directed processive motors. Vacant lattice sites are gray, GTP-dimers are pink and GDP-dimers are green, respectively. Dimers with strong contours are in the excited state as explained in the text. {\bf  b:} Summary of the kinetic transitions of lattice dimers. T (D) denote GTP (GDP) dimers and V denotes vacant lattice sites. The subscript E denotes the excited state. The color code is as in (a). {\bf c:} Summary of the kinetic transitions for molecular motors. }
    \label{fig1}
\end{figure*}

We use a simple, albeit robust, kinetic Monte Carlo model to investigate the MT shaft dynamics in the presence of processive molecular motors. Similar types of kinetic Monte Carlo model have been used e.g.~to study the microtubule tip dynamics \cite{vanburen2002estimates,gardner2011rapid,wu2009simulations,coombes2013evolving,margolin2012mechanisms}, the dynamics of disloctions in the MT shaft \cite{schaedel2019lattice}, and the dynamics of motors walking along the lattice \cite{rank2018crowding}. 
%
The model parameters (see Table \ref{table:par}) are comparable to values found in the literature and are adapted to reproduce typical polymerization ($\sim 2\,\muup$m.min$^{-1}$) and depolymerization speeds ($\sim 20\,\muup$m.min$^{-1}$) of the MT tip, the dynamic instability of the MT plus end (cf.~Supplementary Fig.~\ref{sfig1}) and the typical motor speeds and run length for kinesin and yeast dynein motors.
Here, we specifically aim to study the creation and dynamics of dimer vacancies in the presence of motors using a basic set of kinetic transitions previously established and amended by weak motor-lattice interactions. The basic model setup and kinetic transitions are schematically summarized in Fig.~\ref{fig1}.

\paragraph*{MT lattice structure:} 
We model the canonical microtubule lattice (13 protofilaments, 3-start left-handed helix \cite{mandelkow1986surface,chretien1991new}) as a square lattice on the scale of the dimer, i.e. each dimer has two longitudinal and two lateral neighbours as previously introduced \cite{vanburen2002estimates,wu2009simulations,schaedel2019lattice}.
The lattice is periodic in a direction perpendicular to the long axis of the microtubule with an offset of 3/2 lattice sites to reproduce the seam structure (see Fig.~\ref{fig1}a). 
Lattice sites can be either empty or occupied by GTP-bound (T) or GDP-bound (D) dimers. Dimers interact with other dimers on nearest-neighbour lattice sites via attractive interactions, characterized by bond energies $\Delta G_1$ and $\Delta G_2$ for longitudinal and lateral bonds, respectively. We assume that longitudinal bonds of T–T contacts are further stabilized by the energy, $\Delta G^T_1$
\cite{alushin2014high,Molodtsov2005,Igaev2020}.  We assume a lattice anisotropy in the binding energies of $\Delta G_1/\Delta G_2 = 2$, that is, longitudinal contacts are twice as stable as lateral contacts in the GDP lattice. 
%

%

\paragraph*{Basic MT lattice transitions:}

We consider the following transitions: free GTP dimers can polymerize into a lattice structure, bound GTP dimers can depolymerize from the lattice or hydrolyse into GDP dimers; and bound GDP dimers can depolymerize from the lattice (Fig.~\ref{fig1}b). 
GTP-dimers attach to a vacant lattice site with rate constant $k_\mathrm{on}$, if at least one neighboring lattice site is occupied by a dimer. We assume an infinite reservoir of free GTP-tubulin dimers at concentration $c$. 
We do not consider the attachement of GDP-dimers, i.e. for free tubulin the exchance GDP$\rightarrow$GTP is rapid compared to dimer attachment.  
Dimers detach from the lattice with rate constant $k\lrm{off}$ following the principle of detailed balanced (for details see {\bf Methods})
\begin{equation}
k_\mathrm{off}=k^\ast_\mathrm{off}e^{\beta(\Delta G\lrm{b}+\delta-\Delta G^\ast)} \label{eq:koff}
\end{equation}
where $\beta^{-1}=k_\mathrm{B}T$ is the thermal energy, $\Delta G\lrm{b}$ denotes the binding energy (upon transferring a free dimer from the solution into the lattice), $\delta$ denotes a (weak, transient) contribution due to lattice-motor interactions and $\Delta G^\ast$ denotes the binding energy at
the microtubule tip with a GTP cap (i.e. the binding energy of a GTP-dimer with one lateral and one longitudinal GTP dimer neighbor).

GTP dimers are irreversibly hydrolysed into GDP dimers by the rate constant $k\lrm{hy}$ if their hydrolysable $\beta$ subunit is in contact with the $\alpha$ subunit of another dimer, that is, if the longitudinal lattice site in the direction of the microtubule (+)-end is occupied. 
One key assumption, relevant only for the MT shaft, is a steric hindrance for GTP-dimers to integrate or leave the GDP lattice, if all 4 neighboring lattice sites are occupied by dimers and if the two lateral neighbors are GDP-dimers. Here we follow the observation, that the more extended (in the direction of the protofilament) conformation of GTP-dimers compared to GDP-dimers \cite{alushin2014high,peet2018kinesin} prevents the integration or removal of a GTP-dimer in a GDP-lattice environment, comparable to stacked LEGO bricks. This mechanism has no consequences for the dynamics of the microtubule tip, however, it has far-reaching consequences for the microtubule shaft: once a GDP-dimer has left its GDP-lattice environment and created a point defect, this point defect cannot be closed immediately by integrating a free GTP-tubulin dimer from the solution. 
We will investigate the consequences of this effect extensively in the {\bf Results} section.
%

\paragraph*{Transitions of molecular motors:}

The relevant kinetic transitions for motors are summarized in Fig.~\ref{fig1}c. Here we consider two types of motors, fast kinesin like (+)-end directed motors \cite{Visscher1999,Block2007} and more slow yeast dynein like (-)-end directed motors \cite{Reck-Peterson2006,Rao2013}.
Molecular motors can bind to two adjacent unoccupied tubulin dimers along the same protofilament with rate constant $k_+$ and may detach from the microtubule, when bound to two dimers, with rate constant $k_-$ or, when bound to one dimer, with rate constant $k_-\hrm{E}=\theta k_-$ \cite{gramlich2017single}.  
%
Lattice bound motors step along the microtubule in a single direction (i.e. kinesin towards the (+)-end, dyneins towards the (-)-end) with rate constant $k_w$, if the next lattice site in the stepping direction in not occupied by another motor \cite{rank2018crowding}. We only allow motors to step forward if their front head (in the walking direction) is bound to a dimer. 
If a motor is bound to the lattice by a single dimer (i.e. the second head occupies a vacancy) the underlying dimer may detach from the lattice (with $k\lrm{off}$), taking thereby the motor with it.

\paragraph*{Motor-lattice interactions:}
We consider two types of lattice-motor interactions. (i) Two tubulin dimers which are bound to the same motor are not authorized to leave the lattice. (ii) The irreversible motor step along the protofilament transiently "excites" the underlying tubulin lattice, inducing a slighly less stable ("excited") conformation, i.e.
tubulin dimers (situated under the front head of a bound motor) are excited by the motor walk with rate constant $k\lrm{w}$ by the weak energy increment $\delta$ [see Eq.~\eqref{eq:koff}] and relax back to the "ground" state with the rate constant $k\lrm{r}$. The excitation reaction represents the crucial coupling mechanism between the motor walk and the underlying MT lattice.

\section*{Results}

\subsection*{Creation of a point defect}

When a motor walks along a perfect lattice (i.e.~no point defect present along the protofilament), at each step an underlying tubulin dimer is weakly destabilized by the energy penalty $\delta$. In the following we will explore this effect on the initial creation of a point defect and, vice versa, the effect of the existence of a point defect on the motor walk and the consequence for the lattice stability in the vicinity of the defect. 

\begin{figure*}[ht]
    \centering
    \includegraphics[width=0.9\linewidth]{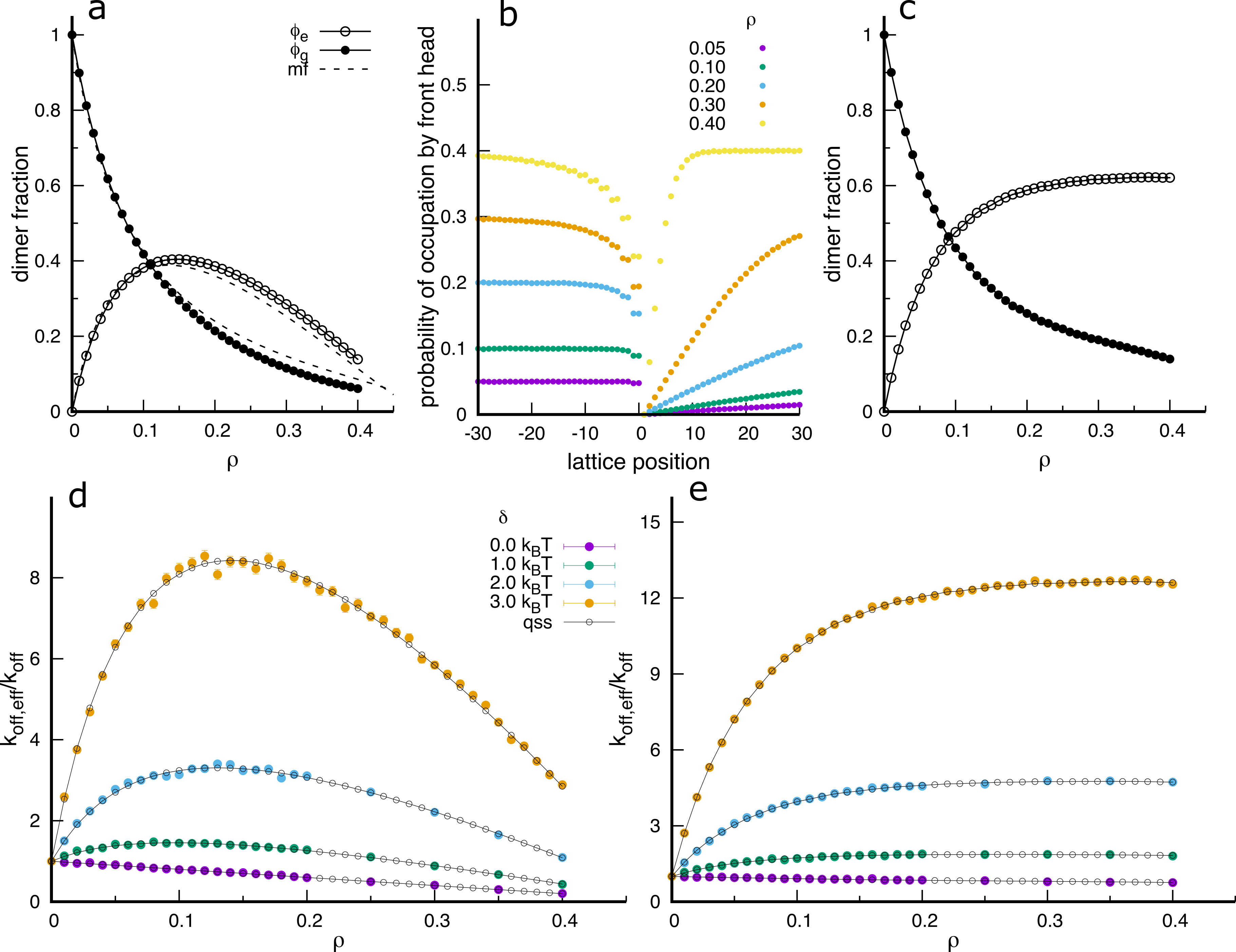}
    \caption{Effect of motor walk (fast motors) on the stability of the MT shaft in the absence of free tubulin. {\bf a:} Quasi steady state fractions of excited ($\phi\lrm{e}$) and ground state dimers ($\phi\lrm{g}$) in the intact lattice depending on the steady state density of motors $\rho$. "mf" indicates the mean field estimate as given in Eqs.~\eqref{phie} and \eqref{phig}. {\bf b:} Quasi steady state probabilities to find a motor front head in the vicinity of a point defect located at the dimer position $i=0$ for various steady state motor densities $\rho$ far away from the defect as indicated in the legend. The motors walk into the positive x-direction. {\bf c:}  Quasi steady state fractions of excited and ground state dimers at a lattice position adjacent to the point defect (upstream). The legend is as in (a). {\bf d:} Effective off-rate constant for the creation of a vacancy (normalized by the off-rate constant of the unperturbed lattice) depending on the motor density $\rho$ for various values of the lattice excitation $\delta$ as indicated in the legend. {\bf e:}  Effective off-rate constant for a tubulin dimer upstream  of a vacancy (normalized by the off-rate constant of the unperturbed lattice) depending on the motor density $\rho$ of the unperturbed lattice for various values of the lattice excitation $\delta$ as indicated in the legend in (d). Remaining parameters are $k\lrm{w}=100\,\tau^{-1}$, $k_-=1\,\tau^{-1}$  and as given in Table \ref{table:par}.} 
    \label{fig2}
\end{figure*}

Fig.~\ref{fig2}a shows the fraction of excited and ground state dimers which can potentially leave the lattice (i.e. they are not sterically blocked by a bound motor)  for fast motors (kinesins) depending on the steady state density of motors on the lattice. The fractions can be roughly estimated using the mean field approach by Rank et al. \cite{rank2018crowding}
\begin{eqnarray}
\phi\lrm{e} & = & (1-2\rho) {J\over J + k\lrm{r}} \label{phie}\\
\phi\lrm{g} & = & (1-2\rho) {k_r\over J + k\lrm{r}} \label{phig}
\end{eqnarray}
where $\rho$ denotes the steady state density of motor front heads and $J=k\lrm{w} \rho (1-2\rho)/(1-\rho)$ denotes the steady state motor flux. The mean field estimates (dashed lines in Fig.~\ref{fig2}a) are slightly off the numerical results which is probably a consequence of the localized appearance of excited dimers and the strong correlations with the motor positions. For low motor densities the fraction of "excited" dimers is increasing at the expense of the fraction of "ground"-state dimers. However, for high motor densities, both fractions of excited and ground-state dimers are decreasing. The behavior reflects the jamming of motors (which reduces excitation of dimers) and the blocking of dimers from leaving the lattice by the bound motors.

If motors encounter a point defect, the steady state motor density is perturbed since motors cannot walk 'over' the point defect and have to detach (see Fig.~\ref{fig2}b). For fast motor detachment at the defect, the motor occupation of dimers in the vicinity of the point defect is lower than in the intact lattice.
Fig.~\ref{fig2}c shows the fraction of excited and ground state dimers which can potentially leave the lattice immediately upstream (w.r.t. the walking direction of motors) to the point defect depending on the steady state density of motors on the lattice. 
In contrast to the intact lattice the fraction of excited dimers is increasing monotonously with the steady state motor density (far away from the defect). 
Here the fast detachment of the motors at the defect guaranties a finite motor flux even at jamming conditions far away from the defect. 
The motors do not affect the lattice immediately downstream of the vacancy (data not shown), since the motor density is here close to zero.

The weak destabilization of a tubulin dimer by the motor walk and the steric blocking of dimers by the presence of motors modify the rate constant $k_\mathrm{off}$ for the dimer to detach from the lattice and to create a vacancy. Since excitation/relaxation processes and the motor walk are fast compared to the residence time of dimers in the full lattice we can estimate the impact of the motor walk on the lattice using a quasi-steady-state assumption
\begin{equation}
    k_\mathrm{off,qss}=k_\mathrm{off}\left[\phi\lrm{e} e^{\beta\delta}+\phi\lrm{g}\right]\,,\label{koff:qss}
\end{equation}
where  $k_\mathrm{off}$ denotes the dimer off-rate constant in the absence of motors. $\phi\lrm{e}$ and $\phi\lrm{g}$ are determined by the motor speed $k\lrm{w}$ and the lattice relaxation rate constant $k\lrm{r}$.   
Fig.~\ref{fig2}d shows the effective rate constant of dimer removal $k_\mathrm{off,qss}/k\lrm{off}$ depending on the steady-state motor density for kinesin motors and a fast lattice relaxation ($k\lrm{w}/k\lrm{r}=10$ implying that at low motor density each motor walks with a trail of 10 excited dimers in its wake) for various values of the motor inflicted tubulin dimer destabilization values $\delta$. 
%
%
Motors, which do not destabilize the lattice ($\delta=0$), have a slightly stabilizing effect, due to sterically blocking the dimers from leaving the lattice. However, a small transient perturbation of the lattice as small as $\delta=3\,k_\mathrm{B}T$ leads to a 10-fold increase in the off-rate constant for moderate motor lattice occupations of $\rho=0.1$ (note that $\rho=0.5$ indicates a lattice completely saturated with motors, each attached with two heads to two adjacent dimers).

Using slower walking dynein motors and maintaining a fast lattice relaxation ($k\lrm{w}/k\lrm{r}=1$) we obtain qualitatively the same results, however higher energy penalties $\delta$ are needed to produce the same effect as fast walking motors, since the density of excited dimers is lower  (see Supplementary Fig.~\ref{sfig2}). Interestingly, for slow motors, the effect of the motor walk on the off-rate constant of dimers located upstream of the vacancy is stronger than on the off-rate constant of dimers located in the intact lattice (cf. Supplementary Figs.~\ref{sfig2}d,e).

Once a vacancy has been created, its effects on the microtubule lattice dynamics are two-fold. i) neighboring dimers are missing a lateral or longitudinal neighbor and therefore dimer detachment is accelerated. ii) The vacancy serves as an obstacle for the motor walk, altering the quasi steady state value of $\phi\lrm{e}$  and $\phi\lrm{g}$ and consequently the effective off-rate constant of dimers upstream of the defect, while the dimers downstream of the defect are depleted of motors.

Therefore, a point defect affects the adjacent dimers in three different ways: lateral dimers experience an unperturbed motor flow and may leave the lattice with an effective off-rate constant shown in Fig.~\ref{fig2}d, longitudinal neighbors upstream of the point defect leave the lattice with an effective rate constant as shown in Fig.~\ref{fig2}e, longitudinal neighbors downstream of the point defect are not affected by the presence of motors. 


\subsection*{Microtubule fracture in the absence of free tubulin}

In the absence of free tubulin a vacancy will lead to the loss of more dimers and a fracture will expand longitudinally and laterally along the shaft, until the MT breaks completely. Due to the lattice anisotropy (longitudinal bonds are assumed to be stronger than lateral bonds) the damage will predominantly expand in a longitudinal direction. 

\begin{figure*}[ht]
    \centering
    \includegraphics[width=0.9\linewidth]{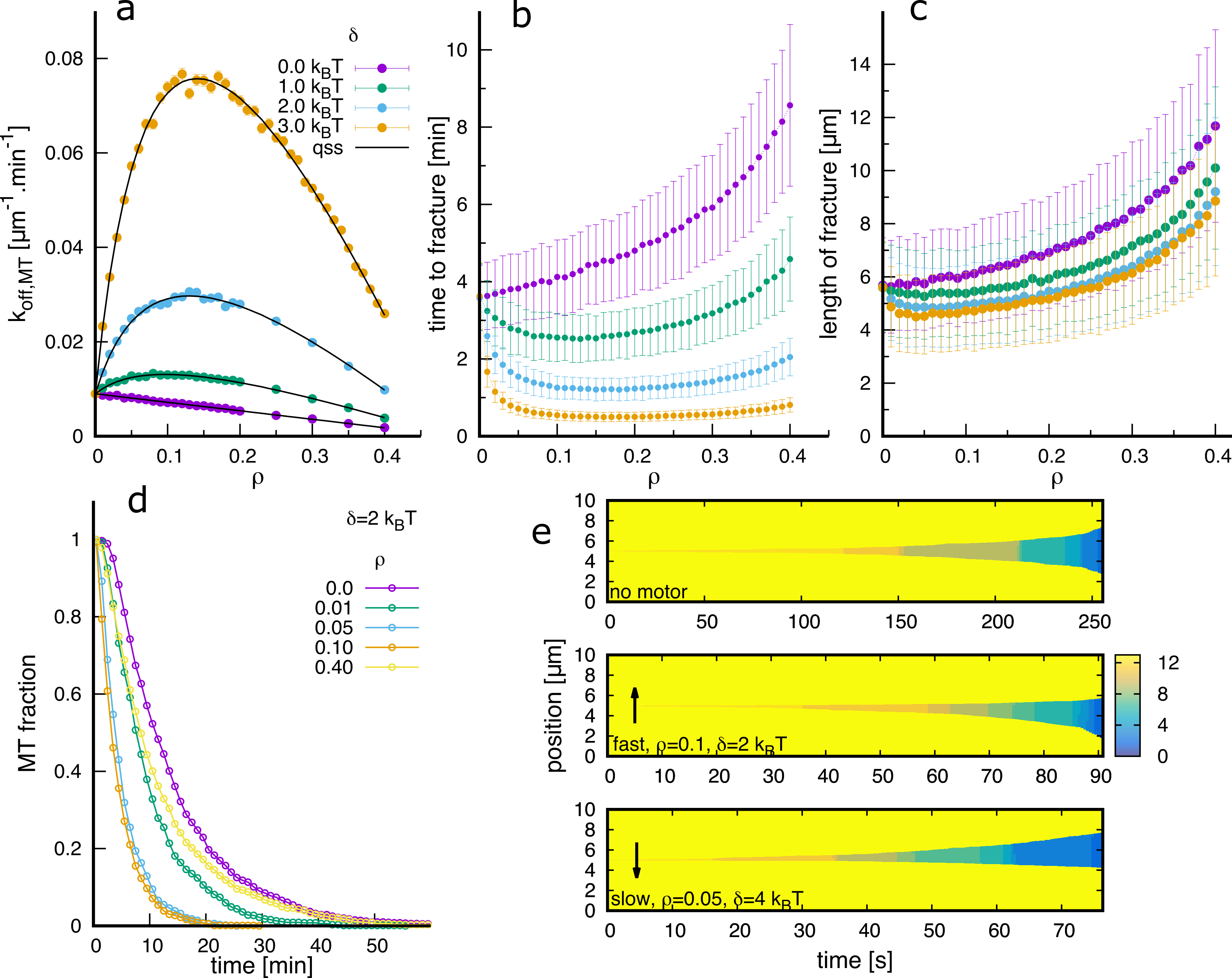}
    \caption{Effect of motor walk on the fracture of the MT shaft. {\bf a:} Effective rate constant for the creation of a vacancy per $\muup$m MT length depending on the motor density $\rho$ for various values of the lattice excitation $\delta$ for fast motors as indicated in the legend. {\bf b,c:}  Time to fracture and length of damaged region at fracture after the creation of a vacancy depending on the motor density $\rho$ for various values of the lattice excitation $\delta$ for fast motors as indicated in the legend in (a). {\bf d:} Survival curves of MTs for various values of the motor density $\rho$ for $\delta=2\,k\lrm{B}T$ for fast motors. {\bf e:}  Kymographs of the fracture process. Simulations were started with a point defect in the center of the MT on protofilament 6 (opposite of the seam). The color code corresponds to the number of intact protofilaments. The direction of motor walk is indicated by the black arrow. Remaining parameters are $k\lrm{w}=100\,\tau^{-1}$, $k_-=1\,\tau^{-1}$  for fast motors,  $k\lrm{w}=10\,\tau^{-1}$, $k_-=0.1\,\tau^{-1}$  for slow motors and as given in Table \ref{table:par}.} 
    \label{fig3}
\end{figure*}

Using a total lattice binding energy of $\Delta G\lrm{b}=-45\,k_\mathrm{B}T$ and a MT length of 10\,$\muup$m, the typical time for the creaction of a vacancy is of the order of 10\,min in the absence of motors and drops for example to 3\,min in the presence of kinesin motors ($\rho=0.15$, $\delta=2\,k_\mathrm{B}T$, see Fig.~\ref{fig3}a).
%
Fig.~\ref{fig3}b shows the time to fracture (after creation of a single vacancy) in the presence of fast motors with various energy penalties $\delta$. In the absence of motors, the time to fracture is about 3.5\,min. At $\delta=0$ motors are stabilizing the lattice and the time to fracture increases compared to the motor-free case. However, a small energy penalty ($\delta=2\,k\lrm{B}T$) decreases the time to fracture significantly at low motor densities. However, the dependence is non-monotonous, i.e. at high motor densities the time to fracture is again increasing.
Fig.~\ref{fig3}c shows the length of the damaged region in the shaft at fracture. In the absence of motors the typical damage size at fracture is about 6\,$\muup$m.
In the presence of motors, at $\delta=0$ the fracture length increases monotonously, i.e. the damage spreads faster in the longitudinal direction than in the lateral direction, compared to the case without motors. A small energy penalty $\delta$ leads at low motor densities to a small decrease of the fracture length; at high motor densities the fracture length increases. Overall, the energy penalty $\delta$ does not much affect the damage size at fracture.
The numerical results (time to fracture, size of damage at fracture) are comparable to experiments (see Fig.~2 in Ref.~\cite{triclin2021self}, Suppl. Fig.~4 in Ref.~\cite{schaedel2019lattice}). 

Typically in experiments, the effect of motors on the MT stability of end-stabilitized MTs in the absence of free tubulin is presented as survival curves, i.e. the fraction of MTs present depending on time. In these experiments, various effects contribute to the MT destruction besides the nucleation of point defects in the intact shaft. For example the loss of the stabilizing cap and the subsequent rapid MT depolymerization is a major cause of MT destruction. For pedagogical reasons,  although the cap-loss is not described in our model, Fig.~\ref{fig3}d shows survival curves of MTs in the presence of fast walking motors with an energy penalty $\delta=2\,k\lrm{B}T$ for various motor densities. At low motor densities the curves shift to the left, i.e. motors lead to a faster destruction of MTs as observed experimentally in Ref.~\cite{triclin2021self}. However, at high motor densities MT have the same stability as in the absence of motors, reflecting the non-monotonous behavior already evident in Fig.~\ref{fig3}b.

Finally, Fig.~\ref{fig3}e shows three kymographs of MT fracture in the absence (top) and presence of motors (center: fast (+)-end directed motors, $\rho=0.1$, $\delta=2\,k\lrm{B}T$, bottom: slow (-)-end directed motors, $\rho=0.05$, $\delta=4\,k\lrm{B}T$). In the absence of motors, the damage spreads symmetrically in the longitudinal direction. Defect growth in the longitudinal direction is faster than in the lateral direction. The defect growth speed in the longitudinal direction increases with the size of damage in the lateral direction. When the last intact protofilament looses a dimer, fracture is complete.
In the presence of (+)-end directed motors the most obvious effect is that the damage spreads faster towards the (-)-end than to the (+)-end, since motors walking towards the point defect are destabilizing the dimers upstream to the existing damage. The downstream side of the damage is not affected by the motors, due to their high processivity. Note also the difference in the time scale, on which fracture occurs; for the chosen example motors accelerate fracture 2-3 times compared to the case without motors. 
In the presence of (-)-end directed motors the damage spreads faster towards the (+)-end than to the (-)-end.
For completeness, Supplementary Fig.~\ref{sfig3} shows the equivalent of Fig.~\ref{fig3}a,b,c for slowly walking (-)-end directed motors. 

In a final set of calculations we have studied the dynamics of a point defect in the presence of free tubulin dimers at high concentration.
 
\subsection*{Vacancy dynamics in the presence of free tubulin dimers}
\label{sec:vac:dyn}

In the absence of any steric constraints in the lattice, a single point defect will be occupied by a newly incorporated dimer with a typical time $\tau\lrm{on}=(k\lrm{on} c)^{-1}=0.05\,$s (for a concentration of free tubulin $c=20\,\muup$M and the $k\lrm{on}$-rate constant given in Table \ref{table:par}). Conversely, a GDP-dimer sitting next to a vacancy in a longitudinal direction will leave the lattice with a typical time $\tau\lrm{off}=[k\lrm{off}^\ast e^{\beta(\Delta G_1+2\Delta G_2-\Delta G^\ast)}]^{-1}=3.3\,$s$\gg \tau\lrm{on}$ (see Table  \ref{table:par}). Therefore, any point defect appearing in the lattice should be closed immediately in the presence of free tubulin, after which the lattice is repaired completely.

\begin{figure*}[ht]
    \centering
    \includegraphics[width=0.9\linewidth]{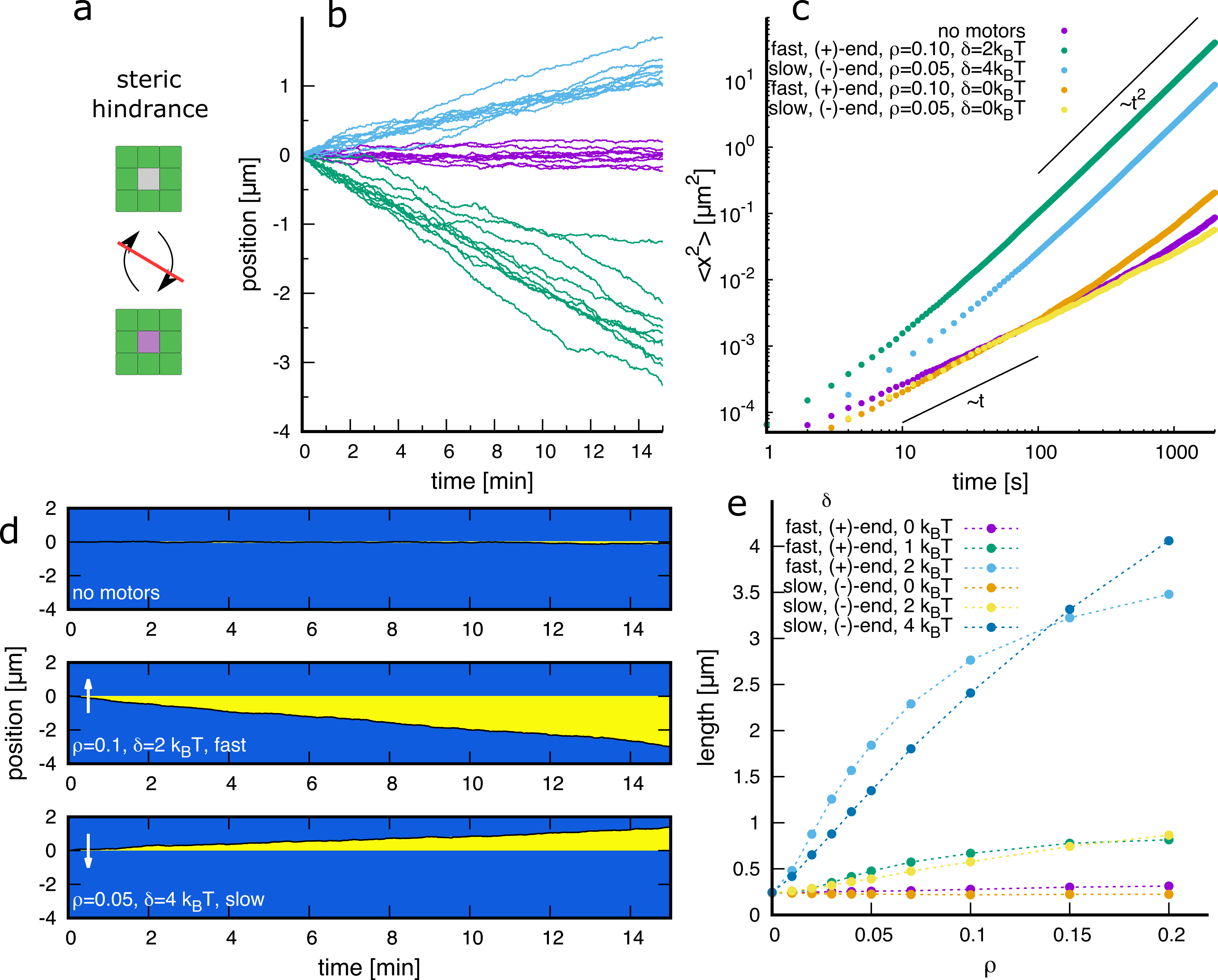}
    \caption{Effect of motor walk on the dynamics of a point defect in the presence of free tubulin. {\bf a:} GTP tubulin dimers (pink) cannot leave or integrate a fully occupied GDP lattice (green) with a single point defect (gray) due to steric hindrance. {\bf b:} Example trajectories of point defects in the absence and presence of motors along a protofilament as indicated in the legend in (c). {\bf c:} Mean squared displacements of point defect trajectories for various motor properties as indicated in the legend. {\bf d:} Examples of kymographs of tubulin exchange (yellow). The current position of the point defect corresponds to the black line. The direction of motor walk is indicated by the white arrow. {\bf e:} Length of tubulin exchange after 15 min depending on the motor density $\rho$ for fast (+)-end directed and slow (-)-end directed motors for various energy penalties $\delta$ as indicated in the legend. Parameters are for fast motors $k\lrm{w}=100\,\tau^{-1}$, $k_-=1\,\tau^{-1}$, for slow motors $k\lrm{w}=10\,\tau^{-1}$, $k_-=0.1\, \tau^{-1}$ and as given in the legends and in Table \ref{table:par}. The concentration of free tubulin is $c=20\,\muup$M.} 
    \label{fig4}
\end{figure*}

\begin{figure*}[ht]
    \centering
    \includegraphics[width=0.9\linewidth]{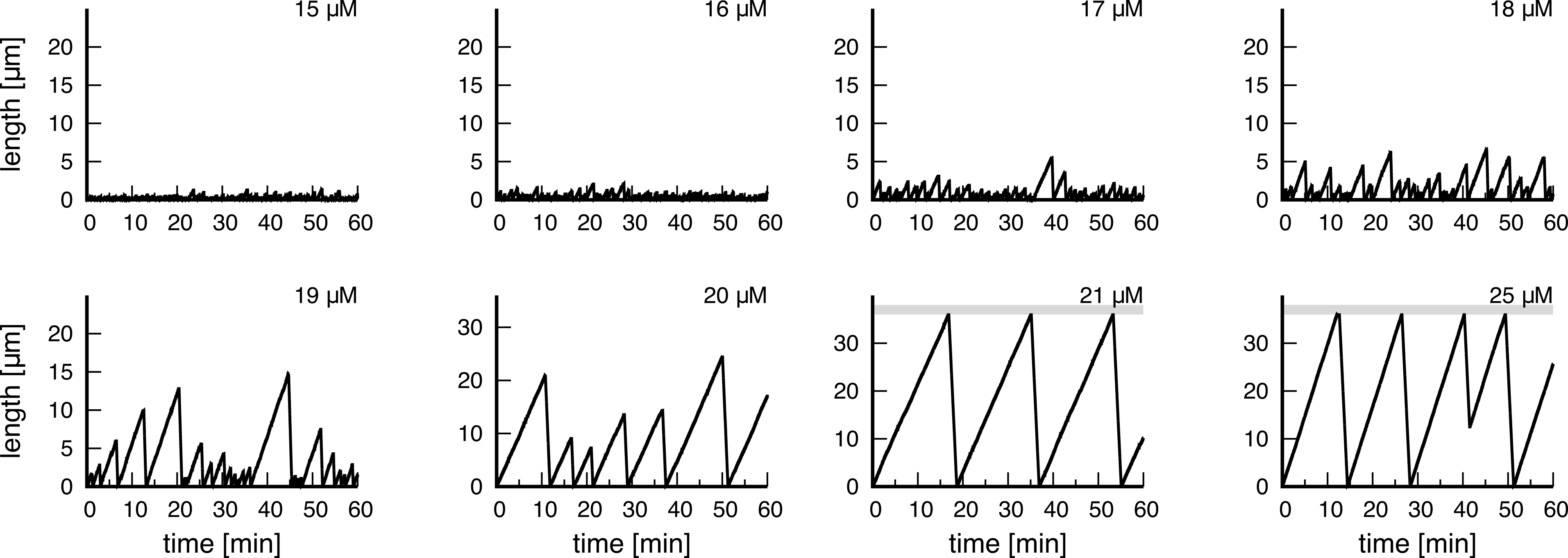}
    \caption{Tip dynamics of the MTs depending on the concentration of free tubulin as indicated in the upper right corner of each graph. The gray bar for 21\,$\muup$M and 25\,$\muup$M denotes the end of the simulation box, which triggers catastrophes. Remaining parameters are as given in Table 1 in the main text.} 
    \label{sfig1}
\end{figure*}

This picture changes completely if we assume that a tubulin dimer experiences a steric hindrance to incorporate the lattice at a single vacancy. Here we have tested the idea, that GTP-dimers cannot integrate a single vacancy due to their extended conformation compared to the GDP-dimer \cite{alushin2014high}. Since GTP-dimer attachment and detachment are reversible, a GTP-dimer cannot leave a GDP lattice environmement (see Fig.~\ref{fig4}a).  This assumption creates a lattice dynamics in the neighborhood of the vacancy, which can potentially survive an extended period of time and lead to localized tubulin exchange experimentally visible in fluorescence microscopy \cite{triclin2021self, andreu2022motor,budaitis2022kinesin}.  
Due to the high anisotropy of the lattice
the vacancy will perform a random walk predominantly along a single protofilament, i.e. up or down the MT axis. In this mechanism, a longitudinal neighbor of a single point defect will leave the lattice, creating a double point defect allowing a free GTP tubulin dimer to attach to either vacant sites. After this cycle of detachment and attachment the lattice has again a single point defect. 

Fig.~\ref{fig4}b shows example trajectories over 15\,min in the absence of motors, and with (+) and (-)-end directed destabilizing motors.
In the absence of motors, the vacancy dynamics is slow and covers only a small distance on the protofilament $\le 0.25\,\muup$m. The trajectories are mainly diffusive, i.e. $\langle x^2\rangle \sim t$ and get slightly super diffusive at long times, due to the stabilizing effects of GTP-contacts in the lattice and the asymmetry in the GTP-hydrolysis (see Fig.~\ref{fig4}b,c). Therefore the vacancy migrates slowly towards the MT (+)-end.

However, in the presence of destabilizing walking motors, the vacancy trajectories are accelerated and become ballistic at long times, i.e. the vacancies are drifting and the mean squared displacement behaves as $\langle x^2\rangle \sim t^2$. The drift direction depends on the walking direction of the motors; (+)-end directed motors induce a drift towards the MT (-)-end, and (-)-end directed motors induce a drift towards the MT (+)-end. 

As an illustration Fig.~\ref{fig4}d shows example kymographs of tubulin exchange, where dimers directly incorporated into the shaft are shown in yellow. 
Fig.~\ref{fig4}e shows the length of incorporation spots depending on the motor concentration and energy penalty $\delta$ for fast (+)-end directed and slow (-)-end directed motors. Without motors at $\rho=0$ the incorporation length is about 0.25\,$\muup$m. Without motor walk induced penalty ($\delta=0$) this length barely changes. However, at a small penalty of $\delta=1\,k\lrm{B}T$ (fast motors) the incorporation length increases  to about 1\,$\muup$m at high motor density ($\rho=0.2$). For $\delta=2\,k\lrm{B}T$ (fast motors) the incorporation length increases strongly with the motor density. Even, at a low motor density ($\rho=0.02$) the incorporation length is about 1\,$\muup$m. Slow motors have a similar effect as fast motors, albeit at higher energy penalties $\delta$.
For comparison, in experiments the incorporation length after 7-40\,min of tubulin incorporation with kinesins (fast motors in our model) are in the range of 1\,$\muup$m \cite{triclin2021self, andreu2022motor,budaitis2022kinesin}.

So far, we have investigated a vacancy dynamics without considering possible repair mechanisms, which could lead to the closure of a point defect. As a weak assumption we could postulate that a lateral GTP-tubulin dimer adjacent to a vacancy allows for the incorporation of a free GTP tubulin dimer, since the GTP-dimer already present in the lattice, sufficiently extends the vacancy site to allow for a GTP-dimer to incorporate. In this picture, a point defect repairs completely, as soon as a vacancy looses a lateral GDP dimer neighbor. In the absence of motors, the typical life-time of a point defect is then given by $\tau\lrm{L}=[k\lrm{off}^\ast e^{\beta(2\Delta G_1+\Delta G_2-\Delta G^\ast)}]^{-1}=100\,$min, well beyond the observation time in an experiment. In the presence of destabilizing walking motors, the life-time may shorten considerably (cf. Fig.~\ref{fig2}d). For example an effective increase in the off-rate constant for a lateral dimer by a factor 3 (Fig.~\ref{fig2}d, $\rho=0.15$, $\delta=2\,k\lrm{B}T$) reduces the life-time of the vacancy to 100/3\,min=33\,min, which is comparable to the experimental time scale. 

\section*{Discussion and Conclusions}

In the present manuscript we have theoretically explored a possible mechanism for MT lattice plasticity in the presence of processive molecular motors. Our basic idea is that the motor walk transiently, locally and weakly destabilizes the underlying MT lattice, which increases the rate of tubulin dimer loss from the shaft lattice. Assuming a sterical hindrance for GTP tubulin dimers to integrate a single vacancy in the GDP lattice induces a lattice dynamics at the vacancy which is accelerated by molecular motors. Furthermore, the vacancy dynamics switches from diffusive to ballistic in the presence of motors whereby the direction of motion depends on the direction of the motor walk.

The proposed model mechanism matches qualitatively and quantitively fracture experiments (fracture size and time to fracture) of end-stabilized MTs in the absence of free tubulin dimers. It is also consistent with experiments on end-stabilized MTs in the presence of free tubulin dimers, which show an increase in the frequency of free dimer incorporation spots with typical sizes in the range of 1\,$\muup$m. 

Frequencies of incorporation spots have been measured by two different groups and vary considerably. Andreu-Carb\'o et al. \cite{andreu2022motor} measured frequencies of 0.05\,$\muup$m$^{-1}$ and  0.2\,$\muup$m$^{-1}$ in the absence of motors and in the presence of kinesins (5\,nM) after 15\,min of free tubulin incorporation, which is consistent with defect nucleation rates shown in Fig.~\ref{fig3}a. Triclin et al. \cite{triclin2021self} measured frequencies of 0.017\,$\muup$m$^{-1}$ and  0.05\,$\muup$m$^{-1}$ in the absence of motors and in the presence of kinesins (10\,nM) after 40\,min of free tubulin incorporation, i.e. the frequency of incorporation is by a factor of 3-4 lower for an incorporation time which is about a factor 3 longer compared to Ref.~\cite{andreu2022motor}. While the absolute frequency of incorporation may depend on the sensitivity of the experimental setup, both groups find an increase in the incorporation frequency in the presence of motors by a factor of 3 to 4 compared to the control experiment, which corresponds in our model to a motor penalty of about 2\,$k\lrm{B}T$ or slightly above for low kinesin densities on the MT.

%
%

A recent paper by Thery \& Blanchoin \cite{thery2021microtubule} speculates about possible mechanisms of interaction between the processive motor walk and the underlying MT lattice. They juxtapose two different concepts: the motor facilitates the lattice dynamics at dislocations (e.g. changes in protofilament numbers)
(termed the "pickpocket" and "burglar" concept, respectively) as opposed to the idea that the motor weakly destabilizes the perfect lattice in its wake (termed the "roadrunner"). In the first case, the motor walk acts on existing defects, which were created during polymerization. In the latter case, the motor continuously and weakly perturbs the lattice which leads to the nucleation of vacancies.

Here we have investigated the "roadrunner"-concept where MT shaft plasticity is enhanced as a collective motor effect; the motor walk weakly destabilizes the MT lattice on an energy scale of a few $k\lrm{B}T$ and facilitates detachment of dimers from the lattice. The dynamics of an existing vacancy is greatly accelerated upstream of the motor current, whereas downstream of the vacancy the lattice is devoid of motors, leading to a vacancy drift in a direction opposite of the walking direction.  Both concepts, i.e. motors act on existing defects and nucleate new defects, are not mutually exclusive and could act in an additive manner. However, the roadrunner concept ({\sl de-novo} nucleation) would increase the number of sites of lattice plasticity depending on the motor density present on the MT \cite{andreu2022motor}, offering a true mechanism of MT regulation.

It has been suggested in Ref.~\cite{kuo2022force}, that the walk of single kinesins is sufficient to directly remove dimers from the MT shaft as a rare event. 
Indeed, the experiments in Ref.~\cite{kuo2022force} show that the cooperative action of several kinesins may be able to remove a tubulin dimer from the lattice by pulling on the dimer via a flexible tether, although a direct proof of dimer removal is missing. Within this concept of direct dimer removal, the motor walk facilitates vacancy nucleation, as an extremely rare event. It is not clear how the motors affect the lattice in the vicinity of the vacancy, since this would require that a rare lattice destabilization event occurs at an existing vacancy. A direct dimer removal offers therefore no straightforward explanation for $\muup$m- sized tubulin incorporation spots, which involve an exchange of several 10 to 100 dimers length.
%
%
However, the direct detachment of tubulin dimers by the kinesin walk as a rare event can be treated potentially within the same kinetic Monte Carlo framework we have used here. To that end, only an almost vanishing small fraction of motor steps leads to a strong transient lattice destabilization (limited by the free energy of ATP hydrolysis).  An investigation of this "single-molecule" mechanism with w.r.t. to MT fracture and free tubulin incorporation is part of future work in an attempt to oppose the two mechanisms of motor-lattice interactions.

It has been speculated, that GTP-islands in the GDP shaft may serve as rescue sites for rapidly depolymerizing MTs and thus an increased tubulin turnover in the shaft may entail an increased MT stability. Indeed, it has been shown in vivo and in vitro that an increased frequency of tubulin exchange sites correlates with a higher rescue frequency \cite{aumeier2016selfrepair,andreu2022motor}.  
Increasing the stability of dynamic MTs by increasing the MT shaft plasticity constitutes a completely novel and unexplored mechanism of MT regulation, of which a mechanistic picture is completely missing. Our proposed mechanism for MT-motor interactions could serve as an important cornerstone in plasticity induced MT regulatory mechanisms.  

\section*{Methods}
\subsection*{Model details and choice of parameters}

Kinetic Monte Carlo simulations were performed using a rejection-free random-selection method \cite{lukkien1998efficient}.
and using custom written codes in C and python. Statistical analysis was either performed using C, python, or R. Unless stated otherwise, kinetic Monte Carlo simulations we performed using the parameters shown in Table \ref{table:par}.

\paragraph*{Details of the seam structure:}
Individual lattice sites on the square lattice are identified by a doublet of integers (i,j).
Lattice sites at the seam have 2 nearest lateral ‘half ’ neighbours across the seam, that is, dimers at the seam in Fig.~\ref{fig1}a with the doublet (1,j) are in contact with dimers (13,j + 2) and (13,j + 1) at the opposite site of the seam and dimers at the right seam with doublet (13,j) are in contact with dimers (1,j – 1) and (1,j – 2) for a 13$_3$ protofilament lattice. 

\paragraph*{Principle of detailed balance for reversible reactions:}
For the passive process of lattice polymerization and depolymerization, the principle of detailed balance has to hold. Therefore, on and off rate constants, $k_\mathrm{on}$ and $k_\mathrm{off}$, must be coupled by the relation \cite{degroot1984non}
\begin{equation}
    {k_\mathrm{off}\over k_\mathrm{on}c_0}=e^{\beta\Delta G}\,,\label{koff:kon}
\end{equation}
where $\beta^{-1}=k_\mathrm{B}T$, $c_0$ denotes the standard concentration of free tubulin in solution, that is,
1\,M by convention, and $\Delta G$ denotes the change in free energy upon transferring a free dimer from the solution into the lattice. Note that $c_0$ in Eq. \eqref{koff:kon} is not the actual concentration of the free tubulin in solution, but the standard concentration and originates from the concentration dependence of the chemical potential,
that is, $k_BT \ln(c/c_0)$, where $c$ denotes the actual concentration of free tubulin in solution. 
$\Delta G = \Delta G\lrm{b} + \Delta G\lrm{e} + \delta$  contains contributions from binding of the dimer to nearest neighbours $\Delta G\lrm{b}$, the loss of entropy due to immobilization of the free dimer in the lattice $\Delta G\lrm{e}$, and an additional (weak) transient contribution $\delta$ due to the stepping of processive motors. For practical reasons, we rewrite Eq.~\eqref{koff:kon} into
\begin{equation}
 k_\mathrm{off}=k_\mathrm{on} c_0 e^{\beta\Delta G}=k^\ast_\mathrm{off}e^{\beta(\Delta G\lrm{b}+\delta-\Delta G^\ast)} \label{sup:eq:koff}
\end{equation}
with $k_\mathrm{off}^\ast=k_\mathrm{on}c_0 e^{\beta(\Delta G\lrm{e}+\Delta G^\ast)}$. $\Delta G^\ast=\Delta G_1+\Delta G_1\hrm{T}+\Delta G_2$ denotes the binding energy at
the microtubule tip with a GTP cap (i.e. the binding energy of a GTP-dimer with 1 lateral and 1 longitudinal GTP dimer neighbor). 

\paragraph*{Justification for the choice of $\Delta G_1^\mathrm{T}$:}
Note that the stabilization of longitudinal T–T contacts is a major difference to the model by vanBuren et al. \cite{vanburen2002estimates}, but permits to capture the dynamic instability without further assumptions. The stabilization of only lateral T–T contacts is not sufficient to induce a dynamic instability with sufficiently long phases of growth and shrinkage. 

\paragraph{Motor dynamics:} Typically, simulations were carried out for a given quasi steady state motor density $\rho$. The motor (front head) density is related to the motor on-rate costant $k_+$ and the motor off-rate constant $k_-$ \cite{rank2018crowding} by
\begin{equation}
0=k_+{(1-2\rho)^2\over 1-\rho}-k_-\rho\,. \label{eq:rank:density}
\end{equation} 

\paragraph*{Motor-lattice interactions in the quasi-steady state approach:}

The computational effort of the kinetic Monte Carlo model presented in section \ref{sec:model} increases rapidly with the employed motor density since the system dynamics is driven by two different timescales; on the one hand the tubulin dynamics ($k^{-1}_{\mathrm{off}}(\Delta G^\ast) = 1$\,s) is slow and, on the other hand, the motor dynamics ($k^{-1}_w = 0.1-0.01$\,s) is one to two orders of magnitude faster. It is then a prerequisite to optimize the kMC algorithm to minimize the computational time. One can either reduce the total number of reactions needed to simulate the system (ie. reduce the complexity of the model) or increase the number of reactions per second (ie. optimize the script algorithm). 
Most of the simulations were done with a sharp reduction in the complexity (ie. the number of considered reactions), using a quasi-steady state assumption to capture the effect of the motor walk on the lattice stability [see Eq.~\eqref{koff:qss}]. We checked that the distribution of dimer detachment times $k\lrm{off,qss}^{-1}$ is exponential.
The quasi-steady state assumption was used to calculate MT fracture in the absence of free tubulin.

\begin{table*}[ht]
\begin{tabular}{| l | l | l |p{0.25\hsize}|} 
\hline
\textbf{Parameter}                          & \textbf{Notation} & \textbf{Value}       & \textbf{Remark}                                                                                                                                                         \\ 
\Xhline{2\arrayrulewidth}
longitudinal binding energy (T-D, D-D) & $\Delta G_1$ & -15\,$k_\mathrm{B}T$ & great variability in the values found in the literature \cite{vanburen2002estimates,sept2003physical,vanburen2005mechanochemical,gardner2011rapid,ayoub2017,schaedel2019lattice}, lattice anistropy is $G_1/G_2=2$\\
\cline{1-3}
lateral binding energy (T-D, D-D) & $\Delta G_2$ & -7.5\,$k_\mathrm{B}T$ & \\
\cline{1-3}
longitudinal stabilizing binding energy (T-T) & $\Delta G_1^T$ & -6.3\,$k_\mathrm{B}T$ & \\
\hline
GTP hydrolysis rate constant & $k_h$ & 0.6\,$\tau^{-1}$ & comparabable to experimental values \cite{Melki1996} and values used in other theoretical studies \cite{margolin2012mechanisms,coombes2013evolving,schaedel2019lattice}\\
\hline
entropic loss due to dimer immobilization &$\Delta G\lrm{e}$ & 15\,$k\lrm{B}T$ & estimated from $k\lrm{off}^\ast=k\lrm{on}c_0 e^{\Delta G\lrm{e}+\Delta G_1+\Delta G_2+\Delta G_1\hrm{T}}$, estimates in the literature range from $10-20\,k\lrm{B}T$ \cite{vanburen2002estimates,howard2001mechanics,erickson1989co}\\
\hline
reference energy at the MT tip& $\Delta G^\ast$ & 28.8\,$k\lrm{B}T$ & $\Delta G^\ast=\Delta G_1+\Delta G_2+\Delta G_1\hrm{T}$\\
\hline
on-rate constant (GTP-dimer) & $k_\mathrm{on}$ & 1\,$\muup$M$^{-1}\tau^{-1}$ & overall on-rate constant per 13 protofilament MT is $k_\mathrm{on}^\mathrm{MT}=13\,\muup$M$^{-1}\tau^{-1}$\ comparable to other studies \cite{walker1988dynamic,vanburen2002estimates,gardner2011rapid,Melki1996}\\
\hline
off-rate constant & $k_\mathrm{off}^\ast$ & $\tau^{-1}$ & corresponds to the off-rate constant for a GTP-dimer with one lateral and longitudinal GTP-neighbor \\
\hline
motor speed & $k_w$ & 10-100\,$\tau^{-1}$ & comparable to typical speeds for yeast dyneins (10\,dimers/s) or kinesins (100\,dimers/s) \cite{andreasson2015examining, Reck-Peterson2006, Jha2015, budaitis2022kinesin}\\
\hline
motor off-rate constant & $k_-$ & 0.1-1\,$\tau^{-1}$ & $k_-=0.1\,\tau^{-1}$ with $k_w=10\,\tau^{-1}$ and $k_-=1\,\tau^{-1}$ with $k_w=100\,\tau^{-1}$corresponds to a typical run length of 100 dimers, comparable to experiments on dyneins and kinesins \cite{andreasson2015examining, Reck-Peterson2006, Jha2015, budaitis2022kinesin}\\
\hline
motor off-rate constant (MT end) & $k_-\hrm{E}=\theta k_-$ & $\theta=100$ & we assume motors are not end tracking \cite{gramlich2017single}\\
\hline
motor on-rate & $k_+$ & & adapted to obtain a steady state motor density $\rho$ using Eq.~\eqref{eq:rank:density}  \\
\hline
conformational penalty due to motor stepping (per dimer) & $\delta$ & 0-4\,$k_\mathrm{B}T$ & adapted for our model \\
\hline
lattice relaxation time & $k_r$ & $1\,\tau^{-1}$ & adapted for our model\\
\hline
\end{tabular}
\caption{Parameters for the kinetic Monte Carlo simulations unless stated otherwise. T-T (D-D) indicates a contact between GTP-dimers (GDP-dimers), T-D indicates a contact between a GTP and GDP dimer. The time scale is $\tau=1\,$s.}
\label{table:par}
\end{table*}

\section*{Acknowledgements}
This work was supported by the French National Agency for Research (ANR-18-CE13-0001.) The authors thank Sarah Triclin, Manuel Théry, and Laurent Blanchoin for fruitful discussions.
The computations were performed using the Cactus cluster of the CIMENT infrastructure, supported by the Rhône-Alpes region (GRANT CPER07\_13 CIRA). The authors thank Philippe Beys who manages the cluster.\\


\section*{Author contributions}
W.L. and K.J. developed the model, performed numerical simulations and wrote the paper.

\clearpage

\bibliography{2022_mitu_article.bib}

\section*{Supplementary figures}

\begin{figure*}[ht]
    \centering
    \includegraphics[width=0.9\linewidth]{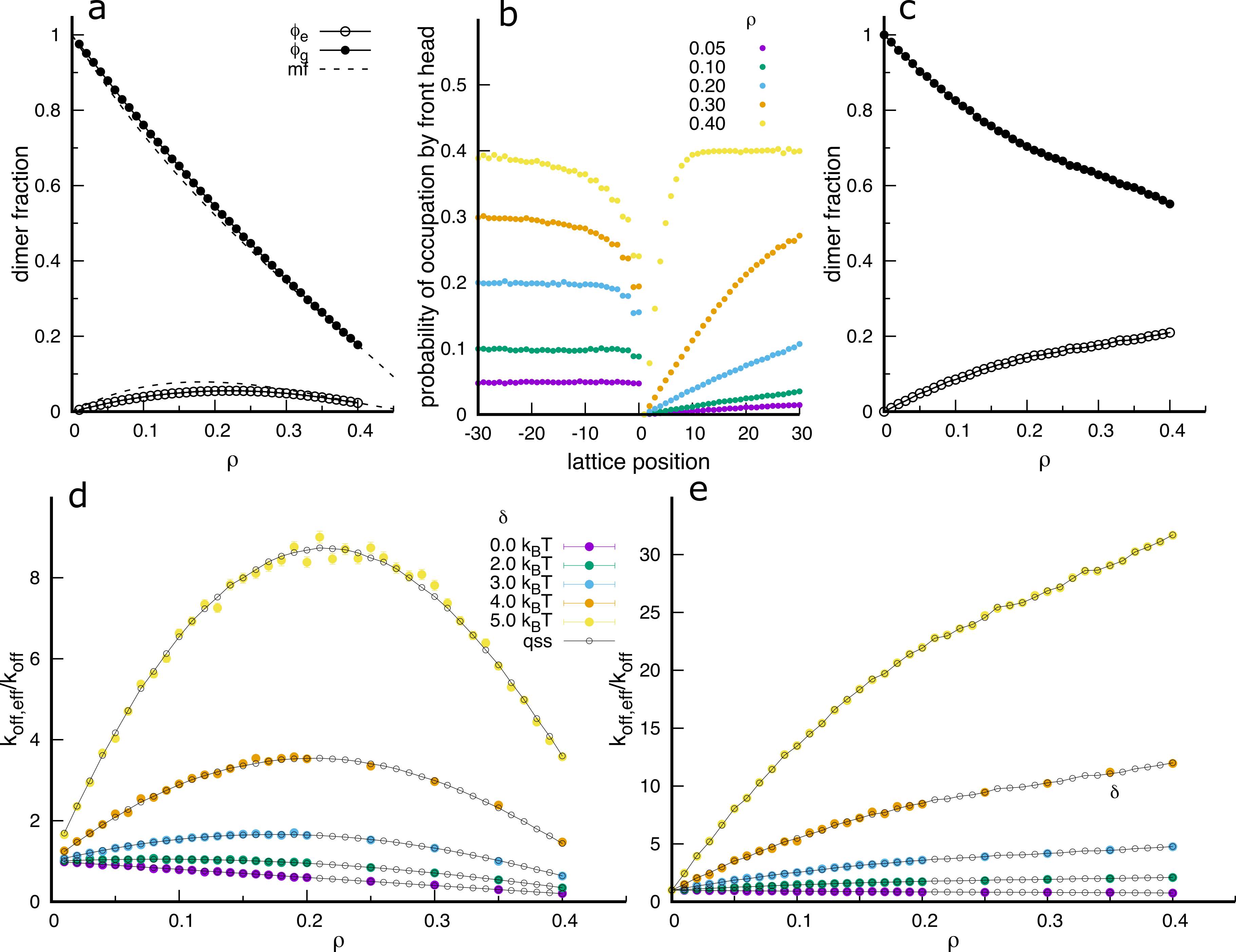}
    \caption{Effect of motor walk on the stability of the MT shaft for slowly walking motors. {\bf a:} Quasi steady state fractions of excited ($\phi\lrm{e}$) and ground state dimers ($\phi\lrm{g}$) in the intact lattice depending on the steady state density of motors $\rho$. "mf" indicates the mean field estimate as given in Eqs.~(2) and (3) in the main text. {\bf b:} Quasi steady state probabilities to find a motor front head in the vicinity of a point defect located at the dimer position $i=0$. The motors walk into the positive x-direction.  {\bf c:}  Quasi steady state fractions of excited and ground state dimers at a lattice position adjacent to the point defect (upstream). The legend is as in (a). {\bf d:} Effective off-rate constant for the creation of a vacancy (normalized by the off-rate constant of the unperturbed lattice) depending on the motor density $\rho$ for various values of the lattice excitation $\delta$ as indicated in the legend. {\bf e:}  Effective off-rate constant for a tubulin dimer upstream  of a vacancy (normalized by the off-rate constant of the unperturbed lattice) depending on the motor density $\rho$ of the unperturbed lattice for various values of the lattice excitation $\delta$ as indicated in the legend in (d). Remaining parameters are $k\lrm{w}=10\,\tau^{-1}$, $k_-=0.1\,\tau^{-1}$  and as given in Table 1 in the main text.} 
    \label{sfig2}
\end{figure*}

\begin{figure*}[ht]
    \centering
    \includegraphics[width=0.9\linewidth]{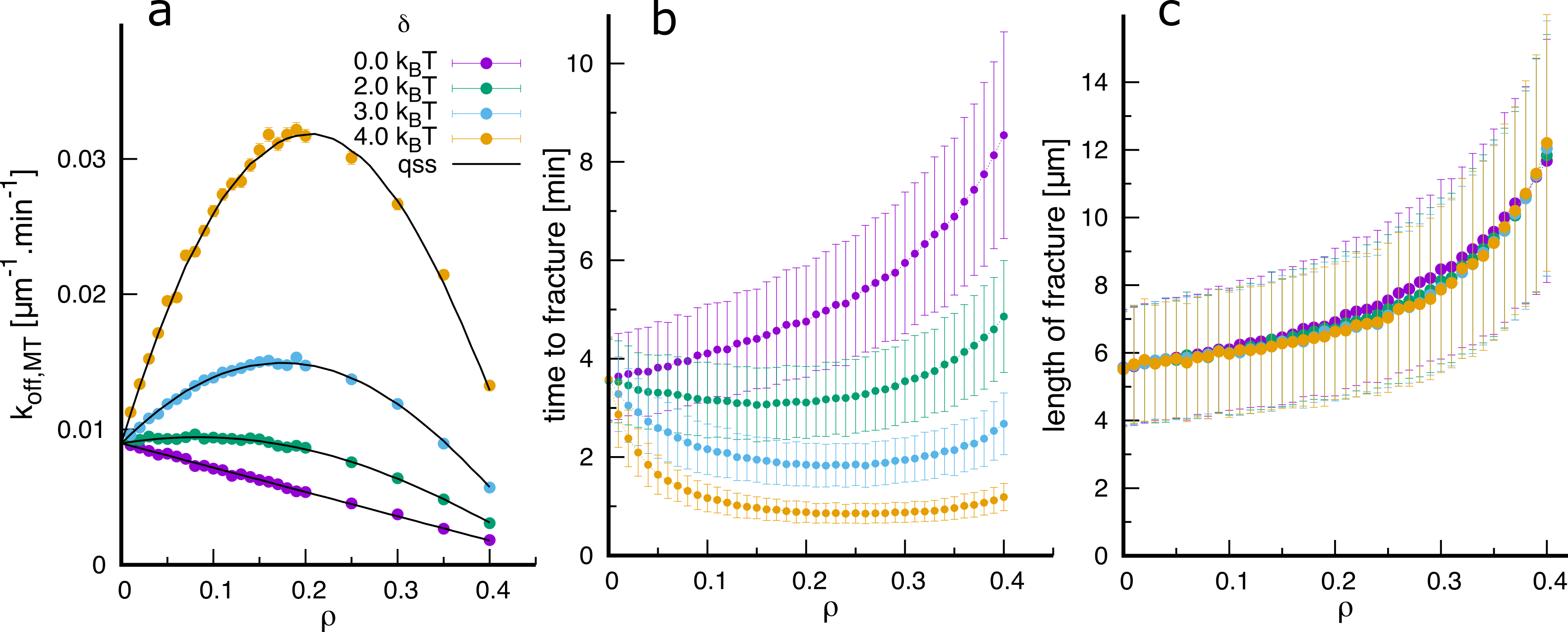}
    \caption{Effect of slowly walking motors on the fracture of the MT shaft in the absence of free tubulin. {\bf a:} Effective rate constant for the creation of a vacancy per $\muup$m MT length depending on the motor density $\rho$ for various values of the lattice excitation $\delta$ as indicated in the legend. {\bf b,c:}  Time to fracture and length of damaged region at fracture after the creation of a vacancy depending on the motor density $\rho$ for various values of the lattice excitation $\delta$ as indicated in the legend in (a). Remaining parameters are $k\lrm{w}=10\,\tau^{-1}$, $k_-=0.1\,\tau^{-1}$  and as given in Table 1 in the main text.} 
    \label{sfig3}
\end{figure*}

\end{document}